\documentclass[pra,aps,twocolumn]{revtex4}
\usepackage{amsmath}
\usepackage{graphicx}

\begin{document}

\title{Creation of effective magnetic fields in optical lattices:\\ The Hofstadter butterfly for cold neutral atoms}

\author{D.~Jaksch$^{1,2}$ and P.~Zoller$^{2}$}

\affiliation{${}^1$Clarendon Laboratory, Department of Physics,
University of Oxford, Oxford, United Kingdom}

\affiliation{${}^2$Institute for Theoretical Physics, University
of Innsbruck, A--6020 Innsbruck, Austria.}

\begin{abstract}
We investigate the dynamics of neutral atoms in a 2D optical
lattice which traps two distinct internal states of the atoms in
different columns. Two Raman lasers are used to coherently
transfer atoms from one internal state to the other, thereby
causing hopping between the different columns. By adjusting the
laser parameters appropriately we can induce a non vanishing phase
of particles moving along a closed path on the lattice. This phase
is proportional to the enclosed area and we thus simulate a
magnetic flux through the lattice. This setup is described by a
Hamiltonian identical to the one for electrons on a lattice
subject to a magnetic field and thus allows us to study this
equivalent situation under very well defined controllable
conditions. We consider the limiting case of huge magnetic fields
-- which is not experimentally accessible for electrons in metals
-- where a fractal band structure, the Hofstadter butterfly,
characterizes the system.
\end{abstract}

\maketitle

\section{Introduction}

The recent experimental progress in manipulating and controlling
trapped neutral atoms in optical lattices by quantum optical means
\cite{bloch02, bloch0201} allows for a number of novel
applications in a variety of different fields like quantum
information processing \cite{brennen99,fermi,jaksch99,jaksch00,
qineu,spinchain}, atom interferometry \cite{spinchain}, and atomic
and molecular physics \cite{atommol,atommol1,atommol2}. One of the
most important features in all of these applications is the large
degree of control by quantum optical techniques over the structure
and the parameters of the Hamiltonian describing the atomic
system. This control allows to realize and deploy a number of
lattice Hamiltonians \cite{Molmer1, Vidal, spinchain} which are
frequently used as toy models for strongly correlated condensed
matter systems and therefore theoretically very well studied.
However, many of the most interesting effects in strongly
correlated 2D systems appear if an external magnetic field is
applied \cite{2dmagn}. Apart from rotating an atomic cloud
\cite{bosereview}, as assumed for example in the study of Laughlin
states with bosonic atoms in Ref. \cite{paredes},  there seems to
be at present no obvious way of implementing lattice Hamiltonians
resembling the effects of magnetic fields with neutral atoms.

\begin{figure}[tbp]
\centering \includegraphics[]{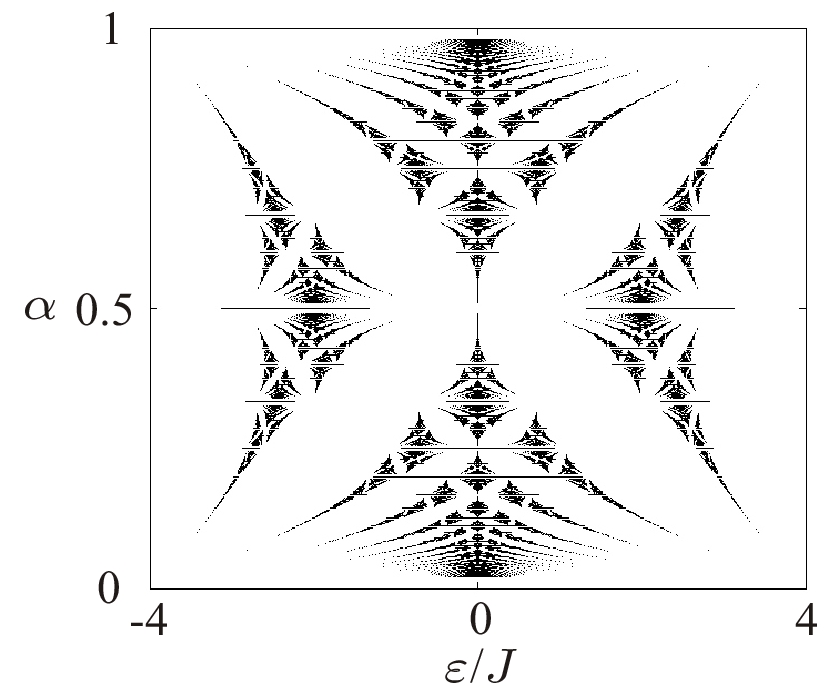} \caption{{\em The Hofstadter
butterfly}: The eigenenergies $\epsilon$ are shown as black dots
for different $\alpha$, and $J$ is the hopping energy as defined
in the text. The most dominant feature is the splitting of the
energy band into $r$ subbands for $\alpha = 1/r$. For further
details of this plot see \cite{Hofstadter}. \label{fig1}}
\end{figure}

In this paper we propose a 2D setup for neutral atoms
which allows to engineer terms in single-band Hubbard Hamiltonians corresponding to
an external magnetic field. An optical lattice provides the
discrete periodic spatial structure and we will use lasers instead
of a magnetic field to induce a phase for particles hopping around
a closed path in the lattice resembling an effective magnetic
field. We will show that the strength of this effective magnetic
field can be varied by laser parameters and be made arbitrarily
large, a situation first theoretically investigated by Hofstadter
\cite{Hofstadter} for electrons. The studies by Hofstadter
predicted the emergence of fractal energy bands $\epsilon$ that
resemble the shape of a butterfly when plotted against the
parameter $\alpha = A B e / 2 \pi \hbar$, where $A$ is the area of
one of the elementary cells of the lattice, $B$ is the strength of
the magnetic field, and $e$ is the charge of the particles (cf.
Fig~\ref{fig1}). The phase $2 \pi \alpha$ is gained by the wave
function of a particle due to the magnetic field when it hops
around a plaquette of the lattice. As shown by Hofstadter the
nature of the energy bands depends crucially on the parameter
$\alpha$. If $\alpha=p/r$ with $p,r$ integers, i.e.~$\alpha$ is a
rational number, the energy spectrum splits into a finite number
of exactly $r$ bands whereas if $\alpha$ is irrational the energy
spectrum breaks up in infinitely many bands and thus the fractal
structure shown in Fig.~\ref{fig1} emerges (for a detailed
discussion on the properties of the Hofstadter butterfly see
\cite{Hofstadter}).

Fractal energy band structures are believed to play an important
role for a number of effects like the quantum Hall effect induced
by magnetic fields in strongly correlated electron systems
\cite{2dmagn, qhall}. Therefore it is desirable to study the
Hofstadter butterfly experimentally under well defined conditions.
However, it turns out that the area of the elementary cells in
metals where fractal energy bands could possibly be seen is so
small that huge magnetic fields would be required to obtain values
of $\alpha$ which are on the order of $1$ \cite{Hofstadter}. Also
in more sophisticated superlattice setups with larger area $A$ it
is experimentally very difficult to obtain direct clear
experimental evidence of the Hofstadter butterfly \cite{hofexp}.

We will consider a 2D system of ultracold atoms trapped in
one layer in the $xy$ plane of a three dimensional optical
lattice. The atoms are in the lowest motional band which can be
achieved e.g.~by loading the optical lattice from a Bose-Einstein
condensate \cite{bloch02,jaksch98}. Hopping along the $z$
direction is turned off completely by the lattice potential.
Different columns of the lattice trap atoms in internal states
$|g\rangle$ ($|e\rangle$) (denoted in Fig.~\ref{fig2}a by open
(closed) circles) \cite{bloch0201,jaksch99}. In addition the
optical lattice is either accelerated along the $x$-axis or an
inhomogeneous static electric field is applied and two Raman
lasers driving transitions between the states $|g\rangle$ and
$|e\rangle$ induce hopping along the $x$-axis while hopping along
the $y$-axis is controlled by the depth of the optical lattice
along this direction. This setup corresponds to applying a
magnetic field with a parameter $\alpha =q \lambda/4 \pi$ where
$q$ is the wave number of the Raman lasers along the $y$ direction
and $\lambda$ the wave length of the lasers creating the optical
lattice. We also note that the atomic setup we are going to
describe here can be used for a large number of other purposes. It
is straightforward to add terms to the system that correspond to
an external electric field. Also the atoms will interact with each
other via collisional contact interactions \cite{jaksch99} and off
site interactions can be engineered by dipolar Rydberg
interactions \cite{jaksch00}. Therefore the setup presented here
can be used for a number of studies related to the behavior of
charged particles in a 2D configuration subject to magnetic and
electric fields and also to study strongly interacting and thus
strongly correlated systems. Furthermore it might be possible to
extend this model to different geometries of optical lattices.

In this work we will concentrate on a possible setup required to
implement the effective magnetic field in an optical lattice. We
will discuss in detail the laser setup which leads to an effective
magnetic flux through the optical lattice, and calculate the
corresponding matrix elements in Sec.~\ref{model}. We also show
that it is possible to reach each point within the Hofstadter
butterfly apart from a negligibly small region around $\alpha=0$
with the proposed setup. In Sec.~\ref{meas} we suggest one
possibility of measuring some of the basic properties of the
Hofstadter butterfly and discuss the limitations on the resolution
for measuring the energy bands. We also give a brief account of
the interaction effects. Finally we conclude with a short outlook
on how the present setup could be extended in
Sec.~\ref{conclusion}. While the focus of the present work is the
derivation of the single-particle terms in the Hubbard Hamiltonian
mimicking a strong magnetic field, we see as one of the main
motivations the extension of strongly correlated many-atom systems
in strong (effective) magnetic fields.

\section{Setup and model} \label{model}

In this section we discuss the experimental setup required to
produce a Hofstadter butterfly for neutral atoms. We first present
the optical lattice setup, then introduce an additional
acceleration or static electric field and finally describe in
detail the additional lasers required for our purpose.

\subsection{Optical lattice}

We consider a three dimensional optical lattice created by
standing wave laser fields which generate a potential for the
atomic motion of the form (we use $\hbar=1$ throughout the paper)
\begin{equation} \label{V}
V(\vec x) = V_{0x} \sin^2(k x) + V_{0y} \sin^2(k y)+V_{0z}
\sin^2(k z),
\end{equation}
with $k=2\pi / \lambda$ the wave-vector of the light and spatial
coordinate ${\bf x}=\{x,y,z\}$. The recoil energy is given by
$E_R=k^2/2M$ with $M$ the mass of the atoms. We assume the lattice
to trap atoms in two different internal hyperfine states
$|e\rangle$ and $|g\rangle$ and the depth of the lattice in $x$-
and $z$-direction to be so large that hopping in these directions
due to kinetic energy is prohibited \cite{jaksch98}. Furthermore
we assume that adjusting the polarization of the lasers which
confine the particles in the $x$-direction allows to place the
potential wells trapping atoms in the different internal states at
distances $\lambda/4$ with respect to each other
\cite{jaksch99,brennen99,bloch0201} as shown in Fig.~\ref{fig1}a.
Therefore the resulting 2D lattice has a lattice constant
(disregarding the internal state) in $x$-direction of
$a_x=\lambda/4$ and in $y$-direction of $a_y=\lambda/2$. We
restrict our analysis to one layer of the optical lattice in the
$xy$ plane since in the following there will neither be hopping
nor interactions between different layers. The dynamics of bosonic
atoms occupying the lowest Bloch band of this optical lattice is
well described by the Bose-Hubbard model (BHM) \cite{jaksch98}
\begin{eqnarray}
&&H_{\rm latt} = \sum_{n,m} J^y \left(a^\dagger_{n,m}
a_{n,m-1}+\rm{h.c.}\right) \nonumber \\
&& \quad + \sum_{n \in \text{even}, m} \omega_{eg} a^\dagger_{n,m}
a_{n,m} +
\sum_{n,m} U_n a^\dagger_{n,m} a^\dagger_{n,m} a_{n,m} a_{n,m} \nonumber \\
&& \quad + U_x \sum_{n,m} a^\dagger_{n,m} a^\dagger_{n+1,m}
a_{n+1,m} a_{n,m}, \label{hamil}
\end{eqnarray}
where $J^y$ is the hopping matrix element for particles to tunnel
between adjacent sites along the $y$-direction. The energy
difference between the two hyperfine states is $\omega_{eg}>0$ and
the operators $a_{n,m}$ ($a^\dagger_{n,m}$) are bosonic
destruction (creation) operators for atoms in the lowest motional
band located at the site which is centered at ${\bf
x}_{n,m}=\{x_n,y_m\}$, where $x_n=n \lambda/4$ and $y_m=m
\lambda/2$. The corresponding mode functions are the localized
Wannier functions $w({\bf x - \bf x}_{n,m})$ \cite{jaksch98} found
by suitable superpositions of the Bloch functions for the lowest
Bloch band of the lattice. Since in $x$-direction the separation
between two neighboring atoms is half the original lattice
constant $\lambda/2$ the overlap of the mode functions of
particles in adjacent lattice sites might lead to significant
nearest neighbor interactions described by $U_x$ whereas we
neglect any other offsite interactions \cite{jaksch98}. The
parameter $U_n$ describes the onsite interaction between two
particles occupying the same site. This onsite interaction may
depend on the column index $n$ because of different internal
states with different scattering lengths occupying different
columns. Since for even [odd] $n$ the operator $a_{n,m}$ describes
atoms in internal states $|e\rangle$, [$|g\rangle$] and the
Wannier functions for sites which are separated by multiples of
the original lattice constant $\lambda/2$ are orthogonal to each
other we find the commutation relations $[a_{n,m},
a^\dagger_{n',m'}] =\delta_{n,n'} \delta_{m,m'}$ with
$\delta_{n,n'}$ the Kronecker delta. The details of the derivation
of the above Hamiltonian can be found in \cite{jaksch98} where
also the definitions for the parameters $U$, $U_x$, and $J^y$ are
given.

\begin{figure}[tbp]
\centering \includegraphics[]{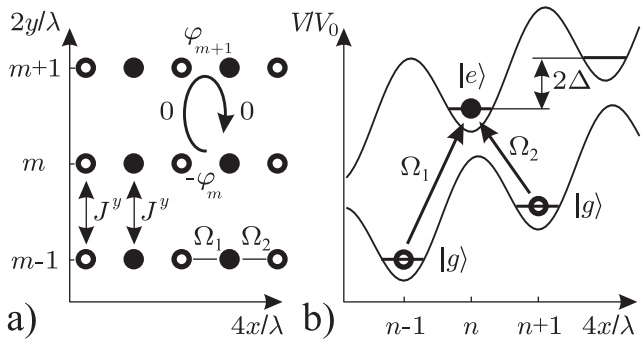} \caption{\textit{Optical
lattice setup}: Open (closed) circles denote atoms in state
$|g\rangle$ ($|e\rangle$). a) Hopping in the $y$-direction is due
to kinetic energy and described by the hopping matrix element
$J^y$ being the same for particles in states $|e\rangle$ and
$|g\rangle$. Along the $x$-direction hopping amplitudes are due to
the additional lasers. b) Trapping potential in $x$-direction.
Adjacent sites are set off by an energy $\Delta$ because of the
acceleration or a static inhomogeneous electric field. The laser
$\Omega_1$ is resonant for transitions $|g\rangle \leftrightarrow
|e\rangle$ while $\Omega_2$ is resonant for transitions between
$|e\rangle \leftrightarrow |g\rangle$ due to the offset of the
lattice sites. Because of the spatial dependence of $\Omega_{1,2}$
atoms hopping around one plaquette get phase shifts of $2 \pi
\alpha=- \varphi_m + 0 + \varphi_{m+1}+ 0$ where $\varphi_m= m q
\lambda/ 4 \pi$ as indicated in a).\label{fig2}}
\end{figure}

\subsection{Acceleration or static electric field}

In addition to the above setup we assume an energy offset of
$\Delta$ between two adjacent sites in $x$-direction as shown in
Fig.~\ref{fig2}b. This can be done by accelerating the optical
lattice along the $x$-axis with a constant acceleration $a_{\rm
acc}$ leading to an additional potential energy term for one atom
of $H_{\rm acc}=M a_{\rm acc} x$. Alternatively, if both of the
internal atomic states $|e\rangle$ and $|g\rangle$ have the same
static polarizability $\mu$ an inhomogeneous static electric field
of the form $E(x) = \delta E x$ can be applied to the optical
lattice yielding a potential energy term $H_{\rm acc}=\mu \delta E
x$. We keep this additional potential energy small compared to the
optical lattice potential and treat $H_{\rm acc}$ as a
perturbation. In second quantization this yields $H_{\rm acc} =
\Delta \sum_{n,m} n a^\dagger_{n,m} a_{n,m}$ where $\Delta= \mu
\delta E \lambda /4$ in the case of an inhomogeneous electric
field and $\Delta =M a_{\rm acc} \lambda /4$ if the lattice is
accelerated. The condition for this perturbative treatment to be
valid is $\Delta \ll \nu_x$ with $\nu_x=\sqrt{4E_{R}V_{0x}}$ the
trapping frequency of the optical lattice in the $x$-direction.

\subsection{Additional lasers}

\begin{figure}[tbp]
\centering \includegraphics[]{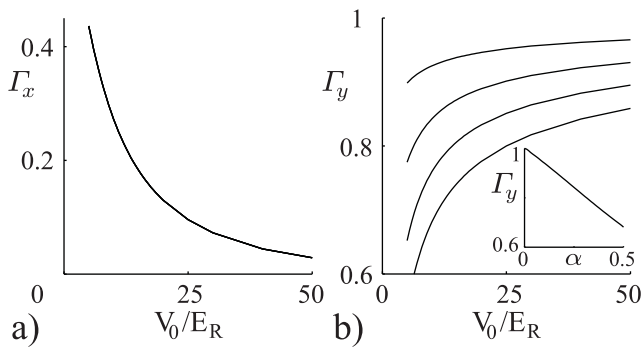} \caption{{\em Matrix
elements:} a) $\Gamma_x$ as a function of the depth of the optical
lattice $V_0/E_R$. b) Matrix element $\Gamma_y(\alpha)$ as a
function of the depth of the optical lattice $V_0/E_R$ for
different values of $\alpha=\{1/8,1/4,3/8,1/2\}$ decreasing with
increasing $\alpha$. The inset shows $\Gamma_y(\alpha)$ against
$\alpha$ at $V_0/E_R=10$.\label{fig3}}
\end{figure}

Finally, we want to induce hopping along the $x$-direction by two
additional lasers driving Raman transitions between the states
$|g\rangle$ and $|e\rangle$. Each of them consists of two running
plane waves chosen to give space dependent Rabi frequencies of the
form
\begin{equation}
\Omega_{1,2}=\Omega e^{\pm i q y}, \label{rabifreq}
\end{equation}
with $\Omega$ the magnitude of the Rabi frequencies, and detunings
$\pm \Delta$. We choose the parameters $\Omega>0$, $\Delta>0$, and
$q>0$. As discussed in detail in Appendix \ref{app1} this can
always be achieved by superimposing two running wave laser beams
incident in the $xy$-plane for $q>(\Delta + \omega_{eg})/c$ with
$c$ the speed of light. We assume the lasers not to excite any
higher lying motional Bloch bands and also no transitions with
detunings of the order of $\Delta$, i.e., $\Omega \ll \Delta \ll
\nu_x$. Then the lasers $\Omega_{1(2)}$ will only drive
transitions $n-1 \leftrightarrow n $ if $n$ is even (odd) and we
can neglect any influence of the nonresonant transitions. We find
the following Hamiltonian describing the effect of the additional
lasers
\begin{eqnarray}
H_{\rm las}&=& \sum_{m,n} \left(\gamma_{n,m} a^\dagger_{n,m}
a_{n-1,m} + {\rm h.c.} \right) \nonumber \\
&&\quad -\Delta \sum_{n,m} n a^\dagger_{n,m} a_{n,m},
\end{eqnarray}
where we have neglected all other terms due to being nonresonant
and defined matrix elements $\gamma_n$ for even $n$
\begin{equation}
\gamma_{n,m}= \frac{1}{2} \int d^3\!x \; {\bf w}^*({\bf x} - {\bf
x}_{n,m}) \Omega_1 {\bf w}({\bf x} - {\bf x}_{n-1,m}),
\end{equation}
and for odd $n$
\begin{equation}
\gamma_{n,m}= \frac{1}{2} \int d^3\!x \; {\bf w}^*({\bf x} - {\bf
x}_{n,m}) \Omega^*_2 {\bf w}({\bf x} - {\bf x}_{n-1,m}).
\end{equation}
For an optical lattice potential of the form Eq.~(\ref{V}) the
Wannier functions can be written as a product of three normalized
Wannier functions, i.e. ${\bf w}({\bf x}) = w(x) w(y) w(z)$
\cite{jaksch98}, and we can write
\begin{equation}
\gamma_{n,m}= \frac{1}{2} e^{2 \pi i \alpha m} \Gamma_y(\alpha)
\Gamma_x,
\end{equation}
where we have defined the matrix elements
\begin{eqnarray}
\Gamma_x &=& \int d\!x \; w^*(x) w(x-\lambda/4), \nonumber \\
\Gamma_y(\alpha) &=& \int d\!y \; w^*(y) \cos(4 \alpha \pi  y /
\lambda)  w(y),
\end{eqnarray}
and $\alpha = q \lambda/4 \pi$. The values of $\Gamma_x$ and
$\Gamma_y(\alpha)$ as a function of the depth of the optical
lattice $V_0$ are shown in Fig.~\ref{fig3}. Both matrix elements
are sufficiently large so that the above inequality $\Omega \ll
\Delta \ll \nu_x$ can be fulfilled if we want to achieve hopping
amplitudes $J^x=\Omega \Gamma_x \Gamma_y /2 \approx J^y$  of the
order of kHz. We simplify $H_{\rm las}$ to find
\begin{eqnarray}
H_{\rm las}&=& \sum_{m,n} \left(J^x e^{2 \pi i \alpha m }
a^\dagger_{n,m} a_{n-1,m} + {\rm h.c.} \right)  \nonumber \\
&&\quad -\Delta \sum_{n,m} n a^\dagger_{n,m} a_{n,m}.
\end{eqnarray}

\subsection{Total Hamiltonian}

The total Hamiltonian describing the configuration shown in
Fig.~\ref{fig2} is given by $H=H_{\rm latt}+H_{\rm acc}+H_{\rm
las}$ and we will for simplicity assume $J^x=J^y=J$. In this work
we will mainly consider small filling factors of the optical
lattice $\bar n \ll 1$ with $\bar n$ the average number of
particles per lattice site and thus only look at one particle
effects neglecting the interaction terms $U_{n}$ and $U_x$. Then
the Hamiltonian can be written as
\begin{equation}
H(\alpha)=J \sum_{n,m} \left(e^{2 \pi i \alpha m} a^\dagger_{n,m}
a_{n+1,m} + a^\dagger_{n,m} a_{n,m+1} + {\rm h.c.}\right).
\label{hofh}
\end{equation}
This Hamiltonian $H(\alpha)$ is equivalent to the Hamiltonian for
electrons with charge $e$ moving on a lattice in an external
magnetic field $B= 2 \pi \alpha /A e$ \cite{Hofstadter}, where
$A=a_x a_y$ is the area of one elementary cell. In the remainder
of this paper we will study the properties of $H$ for neutral
atoms in optical lattices and in particular show that it can be
used to study the whole of the Hofstadter butterfly shown in
Fig.~\ref{fig1}. We note that we have chosen the detunings of the
additional lasers $\Omega_{1,2}$ to exactly cancel the term
$H_{\rm acc}$ arising from the acceleration or electric field. If
these two terms did not cancel the remaining terms resemble a
homogeneous electric field.

\section{Discussion}
\label{meas}

In this section we show how to detect basic properties of the
fractal energy spectrum in an experiment. We also briefly discuss
the effects of fluctuations in the laser parameters, of finite
system sizes, and interactions between the atoms.

\subsection{Measurement}
\label{intpatt}

\begin{figure}[tbp]
\centering \includegraphics[]{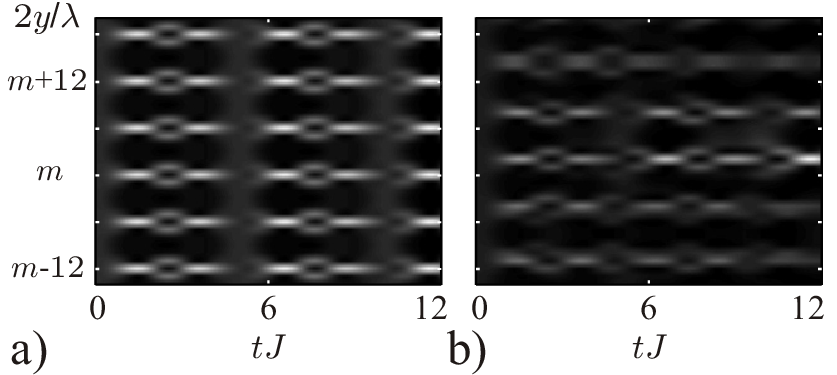} \caption{{\em Particle
density:} $\bar n(y,t)$ (in arbitrary units) as a function of time
$t$ and spatial coordinate $y$. Light (dark) areas indicate a
large (small) particle density. a) For $\alpha=1/6$ a periodic
particle density with a period of exactly $6$ lattice sites
emerges. b) For a value of $\alpha=1/2 \pi$ the periodicity in the
particle density disappears and the visibility of the interference
fringes decreases. \label{fig4}}
\end{figure}

One way of experimentally identifying the number of energy bands
in the case of rational $\alpha$ is to measure the time evolution
of the particle density $\bar n(y,t)$ in the lattice. We assume
the lattice to be loaded from a Bose-Einstein condensate
\cite{bloch02} neglect any interaction between the atoms and
consider only a single atom. The initial wave function of the
system is assumed to be
\begin{equation}
|\Psi(t=0)\rangle =\frac{1}{\cal N} \sum_{n,m} a_{n,m}^\dagger
|\text{vac}\rangle,
\end{equation}
with $|\text{vac}\rangle$ the vacuum state and $\cal N$ a
normalization constant. We then turn on the additional lasers to
simulate a magnetic field and find the density of particles given
by
\begin{equation}
\bar n(y,t)=\langle \Psi(t) |a_{n,m}^\dagger a_{n,m} | \Psi(t)
\rangle,
\end{equation}
to be independent of $x$. If we choose $\alpha=1/r$ a periodic
interference pattern emerges in the time evolution of
$|\Psi(t)\rangle$. Different paths for hopping around in the
lattice interfere with periodic phase relations
(cf.~Fig.~\ref{fig4}a). The periodicity of interference pattern is
determined by $r$ and repeats itself after exactly $r$ lattice
sites as can be seen in Fig.~\ref{fig4}a for $\alpha=1/6$. The
periodicity is destroyed for values of $\alpha$ which are
irrational. An example can be seen in Fig.~\ref{fig4}b where a
value of $\alpha=1/2 \pi$ is chosen which differs by about $5\%$
from $\alpha=1/6$. This little change in $\alpha$ is sufficient to
considerably alter the density of particles in the lattice and to
destroy any periodicity.

\subsection{Parameter fluctuations}

As discussed by Hofstadter \cite{Hofstadter} the maximum number of
energy bands shown in Fig.~\ref{fig1},  which can in principle be
distinguished in an experiment, depends on the fluctuations in the
parameter $\alpha$. In our case these fluctuations are determined
by the frequency stability of the lasers and will thus not be
significant. In an experiment with neutral atoms the resolution of
the energy bands will rather be determined by the size of the
whole sample, i.e. by the number $2 L/\lambda$ where $L$ is the
size of the whole sample, and by the spatial resolution in
measuring interference patterns as described in
Sec.~\ref{intpatt}.

\subsection{Interaction effects}

\begin{figure}[tbp]
\centering \includegraphics[width=8.5cm]{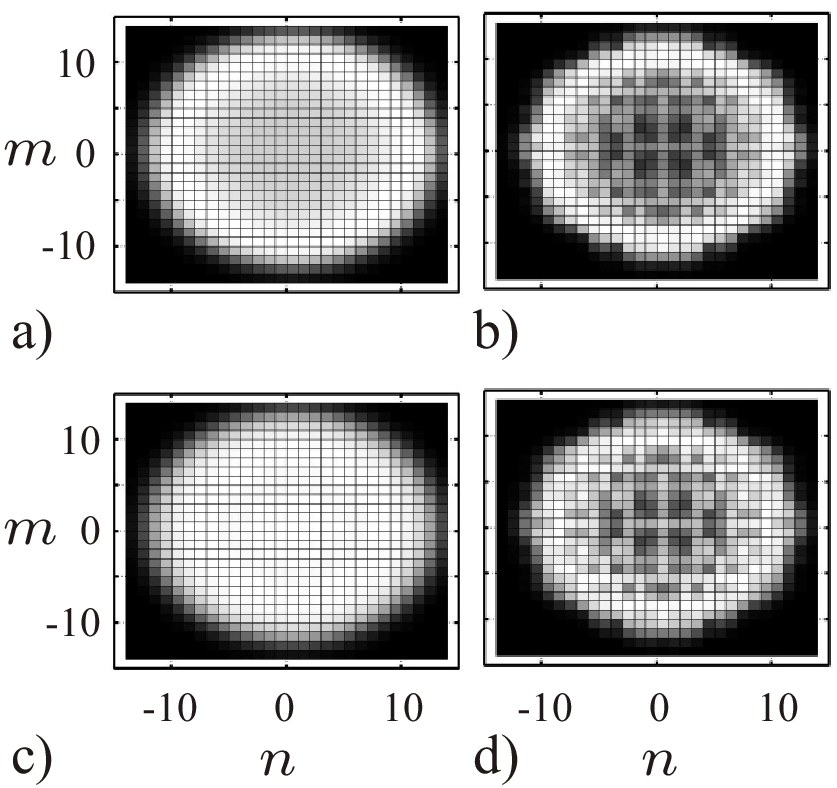} \caption{{\em
Ground state with interactions:} Particle number fluctuations
$\sigma_{n,m}$ [a), b)] and superfluid parameter $|\Phi_{n,m}|$
[c), d)] as a function of $n$ and $m$ for $U=16 J < U_c \approx
5.8 \times 4 J$, $\omega_T= 0.06 J$ and chemical potential
$\mu_c=6 J$. The plots a), c) [b), d)] show the case $\alpha=0$
[$\alpha=1/6$]. Light (dark) areas indicate large (small) values
of the functions (black corresponds to 0 white to 1).
\label{fig5}}
\end{figure}

For large filling $\bar n \geq 1$ the interaction terms in the
Hamiltonian Eq.~\ref{hamil}, especially the onsite interactions
become significant and cannot be neglected. We postpone a detailed
study of these effects to a further publication and only include a
graph of how the ground state in an optical lattice with a
superimposed 2D harmonic trap of trapping frequency $\omega_T$
\begin{equation}
V(x,y)=\frac{M \omega_T^2}{2} \left(x^2 + y^2 \right)
\end{equation}
looks like in the presence of the effective magnetic field. We use
 mean field theory in the form of the Gutzwiller ansatz (as described
in \cite{jaksch98}), and numerically solve for the ground state of
the system for finite $U_n = U$ . In Fig.~\ref{fig5} we plot the
particle number fluctuations $\sigma _{n,m}^{2}=(\langle
\hat{n}_{n,m}^{2}\rangle -\langle \hat{n}_{n,m}\rangle
^{2})/\langle \hat{n}_{n,m}\rangle$ with
$\hat{n}_{n,m}=a^\dagger_{n,m}a_{n,m}$ and the modulus of the
superfluid parameter $\phi _{n,m}=\langle a_{n,m}\rangle$. The
effective magnetic field alters these properties of the ground
state significantly in comparison to the case of $\alpha=0$. The
effective magnetic field leads to a decrease of the particle
number fluctuations and the superfluid density in the center of
the trap which is typical for the onset of a Mott-insulating phase
even for $U<U_c$ with $U_c$ the critical interaction strength for
the transition to the Mott insulator. A similar behavior has
already been found in \cite{monien}. In addition we find from the
numerics that interference effects lead to point like
increase/decraese in $|\phi|$ and $\sigma$ in several of the
lattice sites.

\section{Conclusions}
\label{conclusion}

In conclusion we have shown that quantum optical techniques allow
to implement Hamiltonians often used to model charged particles
subject to an external magnetic field in a lattice. We have shown
that all of the physically interesting parameter regime can be
explored by this setup and proposed one method to measure some of
the most striking features of the fractal energy bands of the
Hofstadter butterfly.

The setup we have investigated possesses a lot of possibilities
for further extensions towards quantum simulations of strongly
correlated systems of charged particles like interacting electrons
moving on a lattice subject to electric and magnetic fields. The
major difference and at the same time one of the most attractive
features of the atomic system is a large degree of control that
can be exerted by quantum optical means. In comparison to
condensed matter systems the Hamiltonian describing the system is
very well known and the parameters appearing can be controlled and
varied over a much wider range than is usually the case for
strongly correlated systems. Also, the time scale over which these
parameters can be changed is short in comparison to decoherence
time scales in the system. This allows to study coherent dynamical
effects that are not easily accessible in most condensed matter
systems. The detailed investigation of such aspects lies beyond
the scope of this paper and will be dealt with in future
publications.

\acknowledgments

P.Z. thanks D. Feder for discussions. This work is supported in
part by the Austrian Science Foundation and EU Networks.

\appendix

\section{Laser configuration}
\label{app1}

We describe the configuration for realizing $\Omega_1$, the
realization of $\Omega_2$ is then straightforward if lasers with
sufficiently different frequencies to avoid interferences between
$\Omega_1$ and $\Omega_2$ are used. Two running laser waves with
Rabi frequencies $\Omega_{e(g)}$ and corresponding laser
frequencies $\omega_{e(g)}$ for driving the transitions $|e\rangle
\leftrightarrow |r\rangle$ ($|g\rangle \leftrightarrow |r\rangle$)
with a large detuning $\delta_r$ are superimposed. We
adiabatically eliminate the auxiliary internal level $|r\rangle$
and the resulting Rabi frequency for the Raman transition between
$|e\rangle$ and $|g\rangle$ is then given by
\begin{equation}
\Omega_1=\frac{\Omega_e \Omega_g}{2 \delta_r} e^{i ({\bf k}_e
-{\bf k}_g) {\bf x}}
\end{equation}
where ${\bf k}_{e(g)} = k_{e(g)} \{\cos(\phi_{e(g)}),\pm
\sin(\phi_{e(g)}),0\}$ is the wave vector of the laser
$\Omega_{e(g)}$. Both lasers are assumed to be incident in the
$xy$ plane at angles $\phi_e$ and $-\phi_g$, respectively. If a
$z$ component of the wave vectors is avoided the experiment can be
done in several identically prepared planes of the lattice
simultaneously which enhances the measurement signal. For a given
$q=k_e \sin(\phi_e)+k_g \sin(\phi_g)$, $\Delta' = (\Delta +
\omega_{eg})/c=k_g-k_e$ fulfilling the requirement that $k_e
\cos(\phi_e) - k_g \cos(\phi_g)=0$ we find
\begin{equation}
\cos\phi_e= \frac{\Gamma}{2 q (k_g - \Delta')}, \qquad \cos\phi_g=
\frac{\Gamma}{2 q k_g},
\end{equation}
where $\Gamma=\sqrt{(q^2-\Delta'^2)(4 k_g^2-q^2-4 k_g
\Delta'+\Delta'^2)}$. These solutions are only physically
meaningful for $\Delta' < q < \sqrt{4 k_g (k_g-\Delta') +
\Delta'^2}$. Since $\Delta' \ll 1/\lambda$ the resulting
limitations on possible values of $q$ do not constrain possible
values of $\alpha$ severely. Only a negligibly small range of
values of $\alpha \approx 0$ will not be realizable due to these
constraints. We note that if we allow the lasers to have a
$z$-component of their wave vector any value of $\alpha$ is
possible.

\end{document}